\documentclass[epj]{webofc}
\usepackage[varg]{txfonts}   
\wocname{EPJ Web of Conferences}
\woctitle{INPC 2013}
\begin{document}
\title{Sensitivity studies for r-process nucleosynthesis in three astrophysical scenarios}

\author{R. Surman\inst{1,2}\fnsep\thanks{\email{surmanr@union.edu}} \and
        M. Mumpower\inst{2} \and
        J. Cass\inst{2} \and
        I. Bentley\inst{3} \and
        A. Aprahamian\inst{2} \and
        G.C. McLaughlin\inst{4}
}

\institute{Department of Physics and Astronomy, Union College, Schenectady, NY 12308 USA 
\and
           Department of Physics, University of Notre Dame, Notre Dame, IN 46556 USA 
\and
           Department of Chemistry and Physics, St Mary's College, Notre Dame, IN 46556 USA
\and
           Department of Physics, North Carolina State University, Raleigh, NC 27695 USA
          }

\abstract{ In rapid neutron capture, or r-process, nucleosynthesis, heavy elements are built 
up via a sequence of neutron captures and beta decays that involves thousands of nuclei far 
from stability. Though we understand the basics of how the r-process proceeds, its 
astrophysical site is still not conclusively known. The nuclear network simulations we use to 
test potential astrophysical scenarios require nuclear physics data (masses, beta decay 
lifetimes, neutron capture rates, fission probabilities) for all of the nuclei on the 
neutron-rich side of the nuclear chart, from the valley of stability to the neutron drip line.  
Here we discuss recent sensitivity studies that aim to determine which individual pieces of 
nuclear data are the most crucial for r-process calculations.  We consider three types of 
astrophysical scenarios: a traditional hot r-process, a cold r-process in which the temperature 
and density drop rapidly, and a neutron star merger trajectory.
}
\maketitle
\section{Introduction}
\label{intro}

One of the major open questions in nuclear astrophysics is the site of the formation of the heaviest 
elements in the r-process of nucleosynthesis (for a review, see e.g., \cite{arn07}).  The r-process 
proceeds via rapid neutron captures, so the challenge is to identify a site that has the requisite 
neutron-to-seed ratio to form nuclei up through $A>200$ \cite{cow91}.  Galactic chemical evolution 
studies favor supernovae \cite{Arg04}.  However, modern simulations of the most promising site within 
a supernova---the neutrino-driven wind off the newly-formed protoneutron star---do not obtain an 
r-process \cite{Arc07,Wan09,Fis10,Hud10}, though uncertainties in the neutrino physics and 
hydrodynamics remain (e.g., \cite{Dua11,Rob12}).  Modern simulations of neutron star mergers, on the 
other hand, show robustly neutron-rich outflows with vigorous r-processing \cite{Gor11,Kor12}, and 
there may even be a hint of the radioactive decay of r-process material observed in a merger event 
\cite{Ber13}.  Even so, it is difficult to understand observations of r-process nuclei in very old 
stars \cite{sne08} if mergers are the sole r-process site.

Simulations of r-process nucleosynthesis that investigate these astrophysical scenarios rely on 
nuclear data such as masses, beta decay rates, and neutron capture rates for thousands of nuclei on 
the neutron-rich side of stability.  Experimental information is only available for a handful of the 
needed masses and even fewer of the necessary beta decay rates, though the situation is improving 
dramatically as the next generation of radioactive beam facilities begin to take data.  We have 
developed a program of r-process sensitivity studies to determine which of the thousands of pieces of 
nuclear data needed should be the targets of new experimental campaigns.  In these sensitivity 
studies, a baseline r-process simulation is chosen, then each piece of nuclear data is varied 
individually and the simulation re-run.  Then the pieces of nuclear data with the greatest impact on 
the final r-process abundance pattern are identified and the mechanisms of their influence are 
determined.  So far sensitivity studies have been performed for neutron capture rates 
\cite{beu08,sur09,sur11,mum12}, nuclear masses \cite{bre12,sur13}, and beta decay rates 
\cite{cas12,sur13}.

Here we review the notable features of the sensitivity study results for separate nuclear mass, 
beta decay rate, and neutron capture rate sensitivity studies for a main ($120<A<200$) hot 
r-process.  We then compare the results to sensitivity studies performed with different types 
of astrophysical conditions: a neutrino-driven wind cold r-process, where the temperature and 
density drop quickly and equilibrium between captures and photodissociations, 
$(n,\gamma)$-$(\gamma,n)$ equilibrium, holds only briefly, and a low entropy, very neutron-rich 
r-process from a neutron star merger simulation.

\section{Sensitivity studies}
\label{sec-1}

We run our sensitivity studies as described in \cite{sur09,bre12,sur13}.  We use an r-process 
nuclear network code from \cite{sur01}, and we take nuclear masses from \cite{mol95}, beta 
decay rates from \cite{mol03}, and neutron capture rates from \cite{rau00}, wherever 
experimental data is not available.  Fission is included as in \cite{beu08b}.  We chose a set of 
astrophysical conditions and produce a baseline r-process simulation. For each baseline 
trajectory chosen, we then run three separate sensitivity studies: a nuclear mass sensitivity 
study, where we vary individual binding energies by $\pm1$ MeV as in \cite{sur13}, a beta decay 
rate sensitivity study, where we vary individual beta decay rates by a factor of 10 as in 
\cite{cas12}, and a neutron capture rate sensitivity study, where we vary individual capture 
rates by a factor of 100 as in \cite{sur09,mum12}.  Finally we compare the final abundance 
patterns produced with the nuclear data variations to the baseline pattern using the 
sensitivity measure $F$:
\begin{equation}
F=100\times \left\lbrace \sum_{A} |AY_{\mathrm{baseline}}(A)-AY_{\mathrm{increase}}(A)| + \sum_{A} |AY_{\mathrm{baseline}}(A)-AY_{\mathrm{decrease}}(A)| \right\rbrace/2
\end{equation}
where $Y_{\mathrm{baseline}}(A)$ are the final baseline abundances, and $Y_{\mathrm{increase}}(A)$ and 
$Y_{\mathrm{decrease}}(A)$ are final abundances of the simulations where a single piece of nuclear 
data is increased or decreased, respectively.

\begin{figure}[h]
\centering
\includegraphics[width=5.0in,clip]{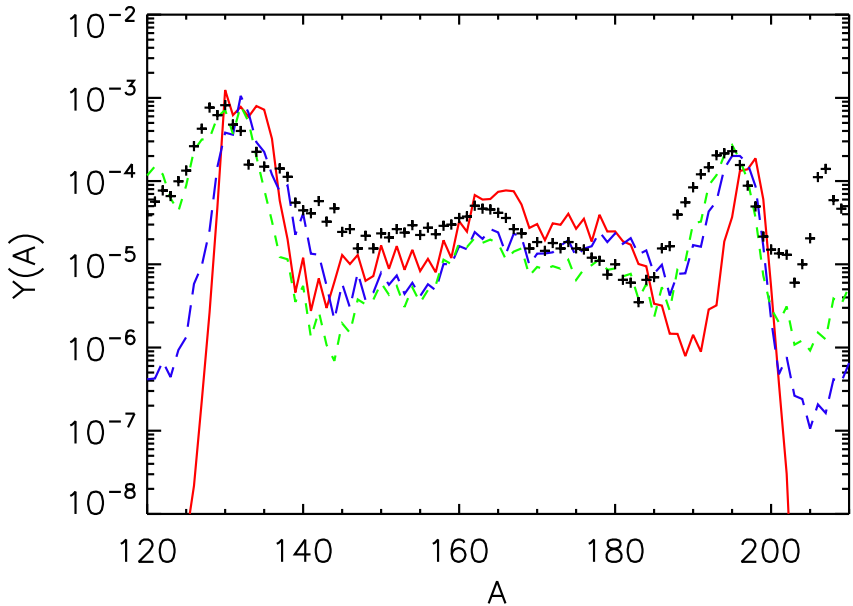}
\caption{Abundances versus mass number for the three baseline r-process simulations described in the 
text: a parameterized wind hot r-process (solid red line), supernova neutrino-driven wind cold 
r-process (long-dashed blue line), and a neutron star merger r-process (short-dashed green line).  The 
scaled solar abundances from \cite{sne08} (crosses) are shown for comparison.}
\label{yva} 
\end{figure}

For this work we chose three distinct astrophysical scenarios for the baseline simulations, with final 
abundance patterns shown in Fig.~\ref{yva}.  The first is a classic hot r-process similar to that 
studied in \cite{beu08,sur09,sur11,mum12,bre12,cas12,sur13}.  A classic hot r-process is characterized 
by an $(n,\gamma)$-$(\gamma,n)$ equilibrium phase during which the main r-process peaks at $A\sim130$ 
and $A\sim195$ are produced, followed by a freezeout phase triggered by the depletion of free 
neutrons.  During the freezeout phase, photodissociations, neutron captures, beta decays, and beta 
delayed neutron emission all compete to set the final abundance pattern and to form the smaller 
abundance peak near $A\sim 162$ called the rare earth peak \cite{sur97,mum12b}.  For this example we 
take a parameterized wind trajectory from \cite{mey02} as implemented in \cite{mum12}, with entropy 
$s/k=50$, dynamical timescale $\tau_{\mathrm{dyn}}=80$ ms, and initial electron fraction 
$Y_{e}=0.250$.

The second astrophysical scenario considered is a cold r-process.  In a cold r-process, the 
temperature and density drop quickly, so $(n,\gamma)$-$(\gamma,n)$ equilibrium is established 
only briefly.  Once the temperature drops, photodissociation effectively turns off, and an a 
new equilibrium is established between neutron captures and beta decays.  For our cold 
r-process example we take a trajectory from the neutrino-driven wind simulations of 
\cite{Arc07} where we artificially reduce the electron fraction to $Y_{e}=0.31$ to produce a 
main r-process.

The final scenario we consider is the ejection of mildly heated, very neutron-rich material 
from the tidal tails of a neutron star merger.  We start with a trajectory from a simulation by 
A. Bauswain and H.-Th.\ Janka, similar to those from \cite{Gor11}.  We then extrapolate it out 
to low temperatures, keeping the entropy constant, as validated by recent merger simulations 
followed to late times \cite{ros13}.  The resulting r-process proceeds in 
$(n,\gamma)$-$(\gamma,n)$ equilibrium as in the classic hot r-process, but here freezeout is 
prompted by the drop in temperature and density rather than an exhaustion of free neutrons as 
in the hot r-process case.  Neutrons are in fact so plentiful that fission recycling occurs.

\section{Results}
\label{sec-2}

\begin{figure}[h!]
\centering
\includegraphics[width=5.0in,clip]{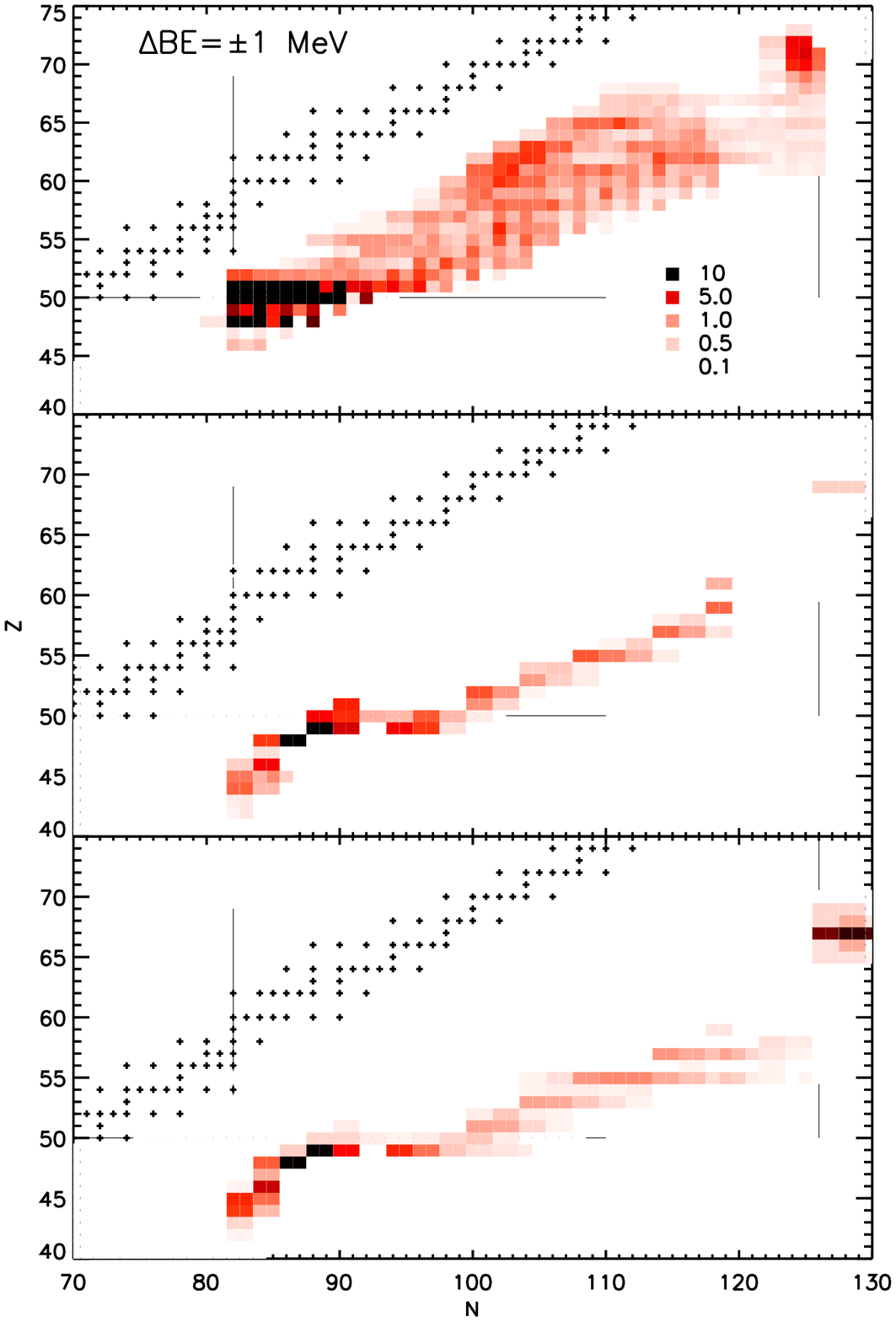}
\caption{Sensitivity measures $F$ for three binding energy sensitivity studies starting from the different 
astrophysical conditions as described in the text: a parameterized wind hot r-process (top panel), 
supernova neutrino-driven wind cold r-process (middle panel), and a neutron star merger r-process (bottom panel).}
\label{fig-1} 
\end{figure}

Fig.~\ref{fig-1} shows the results of the mass/binding energy sensitivity studies for each of the 
three astrophysical scenarios.  As described in \cite{bre12,sur13}, masses appear explicitly in an 
r-process network code only in the calculation of photodissociation rates via detailed balance.  Thus, 
these sensitivity studies point out the importance of nuclear masses in (1) determining abundances 
along the r-process path while $(n,\gamma)$-$(\gamma,n)$ equilibrium holds, and (2) setting the 
individual photodissociation rates that become important when $(n,\gamma)$-$(\gamma,n)$ equilibrium 
fails.  Both of these effects are important for the classic hot r-process, as demonstrated in 
\cite{bre12,sur13} and shown in the top panel of Fig.~\ref{fig-1}.  High sensitivity measures $F$ are 
found for nuclei along the equilibrium r-process path, particularly at the closed shells where the 
main peaks form, as well as for nuclei along the decay paths toward stability, particularly in the 
rare earth region as the rare earth peak forms.  In contrast, the cold r-process study (middle panel 
of Fig.~\ref{fig-1}) and the merger study (bottom panel of Fig.~\ref{fig-1}) show high sensitivities 
only along the equilibrium r-process path.  In the cold r-process, the temperature drops so quickly 
that equilibrium fails before even the $A\sim 195$ peak is formed, and photodissociation plays almost 
no role in the subsequent dynamics.  In the merger case, the main abundance features are indeed formed 
in $(n,\gamma)$-$(\gamma,n)$ equilibrium, but since the onset of freezeout is due to the 
declining temperature photodissociations are not important during the decay back to stability.

\begin{figure}[h!]
\centering
\includegraphics[width=5.0in,clip]{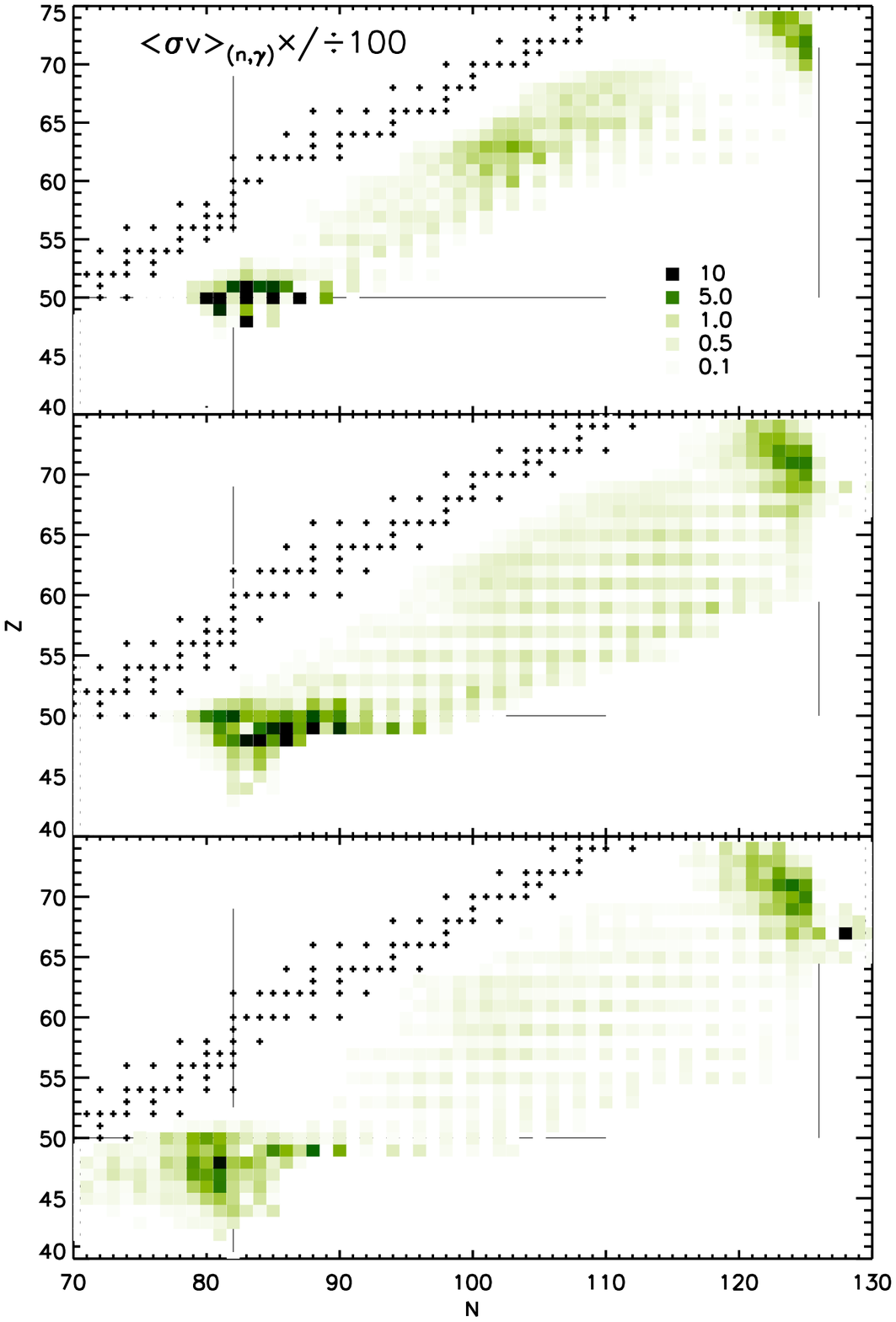}
\caption{Sensitivity measures $F$ for three neutron capture rate sensitivity studies, as in Fig.~\ref{fig-1}.}
\label{fig-2}
\end{figure}

The neutron capture rate sensitivity measures for the same three trajectories are shown in 
Fig.~\ref{fig-2}.  Individual neutron capture rates become important only after 
$(n,\gamma)$-$(\gamma,n)$ equilibrium begins to break down.  Therefore, for a hot r-process, there is 
little sensitivity to neutron capture rates of nuclei along the r-process path during the equilibrium 
phase, as described in \cite{beu08,sur09,mum12} and as shown in the top panel of Fig.~\ref{fig-2}.  
The greatest sensitivities are found for nuclei along the beta decay pathways of the equilibrium 
r-process path nuclei, particularly near the closed shells where the abundances are highest and in the 
rare earth region where they impact the rare earth peak formation mechanism.  In contrast, in a cold 
r-process $(n,\gamma)$-$(\gamma,n)$ equilibrium is quickly replaced by a new equilibrium between beta 
decays and neutron captures.  Thus, the cold r-process example shown in the middle panel of 
Fig.~\ref{fig-2} demonstrates sensitivity to neutron capture rates of nuclei along the early-time 
r-process path, as well as along the beta decay pathways of these nuclei.  The merger example shown in 
the bottom panel of Fig.~\ref{fig-2} shows similar systematics to the hot r-process example, except 
for a much broader range of nuclei since the equilibrium r-process path is so far from stability.

\begin{figure}[h!]
\centering
\includegraphics[width=5.0in,clip]{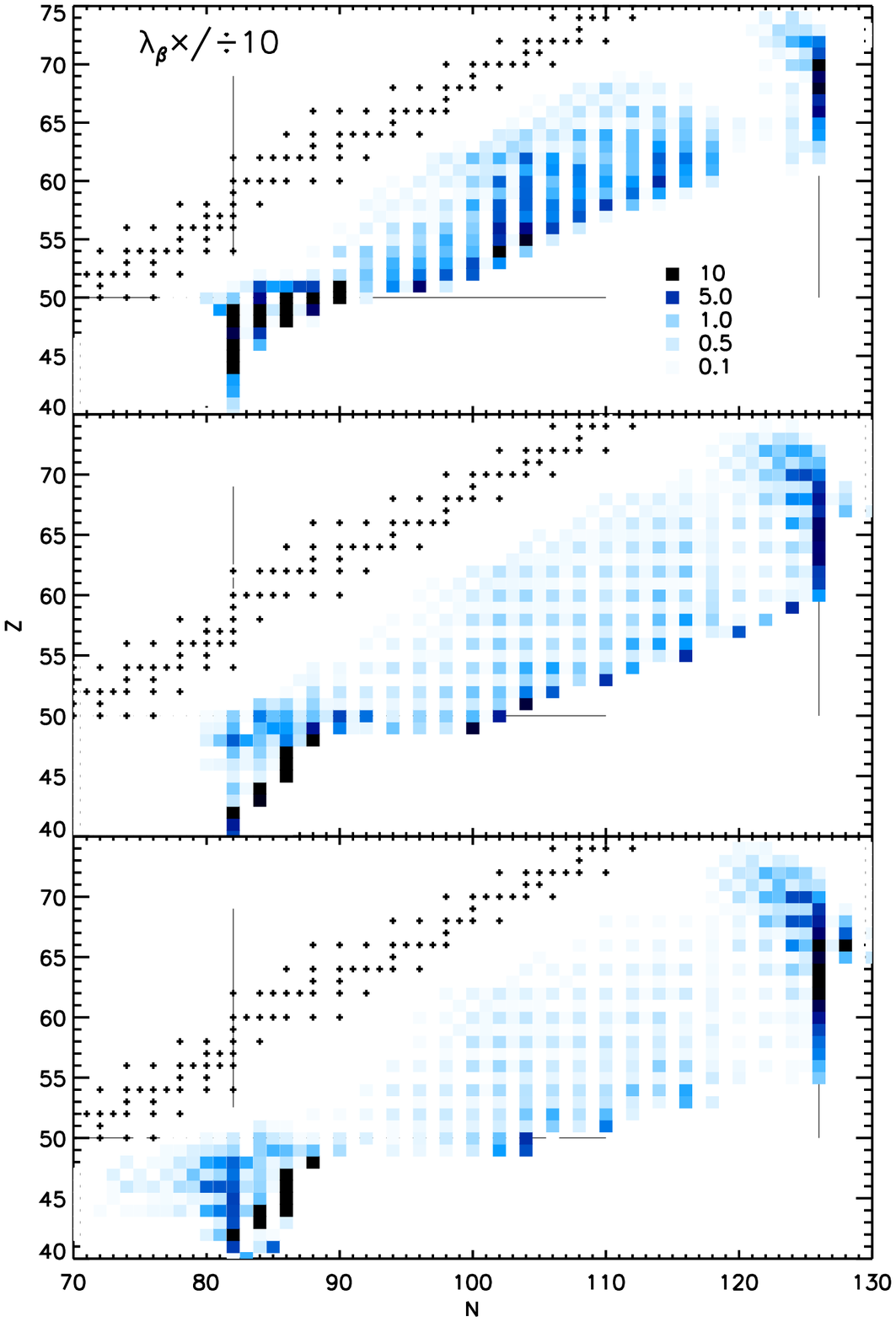}
\caption{Sensitivity measures $F$ for three beta decay rate sensitivity studies, as in Fig.~\ref{fig-1}.}
\label{fig-3}       
\end{figure}

Finally in Fig.~\ref{fig-3} we show the results from the three beta decay rates sensitivity 
studies.  Beta decay rates are of key importance in setting the relative abundances of nuclei 
along the r-process path, and they remain important during the freezeout phase for as long as 
beta decay competes with neutron capture \cite{cas12}. Thus we expect the highest beta decay 
sensitivities for nuclei along the early-time r-process path, with nonzero sensitivities 
extending through nuclei quite close to stability regardless of the astrophysical scenario, as 
shown in all three panels of Fig.~\ref{fig-3}.  An additional notable feature of the beta decay 
sensitivity measures is that they are significantly larger for even-$N$ nuclei than odd-$N$ nuclei.  
This is because odd-$N$ nuclei are much more likely to be depopulated by neutron capture or 
photodissociation than beta decay.

\section{Discussion}
\label{disc}

Studies of r-process nucleosynthesis suffer from large uncertainties in the nuclear physics and 
in the astrophysics.  Here we have examined the sensitivity of the final r-process abundance 
pattern to individual masses, beta decay rates, and neutron capture rates in three very 
different potential astrophysical scenarios: a traditional hot r-process, a supernova 
neutrino-driven wind cold r-process, and a neutron star merger fission cycling r-process.  The 
resulting sensitivities highlight the differences in how the r-process proceeds in each case.  
However, there is one key similarity: the most important pieces of nuclear data in all cases 
are for nuclei near the closed shells and in the rare earth region 10-20 neutrons from 
stability.  We recommend these nuclei be targets of the next generation of experimental 
campaigns.

\section*{Acknowledgements}

This work was supported by the National Science Foundation under grant number PHY1068192 and 
through the Joint Institute for Nuclear Astrophysics grant number PHY0822648, and the 
Department of Energy under contracts DE-FG02-05ER41398 (RS) and DE-FG02-02ER41216 (GCM).

\end{document}